\documentclass[12pt,english]{article}
\usepackage[T1]{fontenc}
\usepackage[latin9]{inputenc}
\usepackage[letterpaper]{geometry}
\usepackage{array}
\usepackage{amsmath}
\usepackage{graphicx}
\usepackage{setspace}
\usepackage{amssymb}

\makeatletter

\providecommand{\tabularnewline}{\\}

\newcommand{\lyxaddress}[1]{
\par {\raggedright #1
\vspace{1.4em}
\noindent\par}
}

\usepackage{subfigure} 

\makeatother

\usepackage{babel}

\begin{document}

\title{Transient Pulse Formation in Jasmonate Signaling Pathway}

\author{Subhasis Banerjee and Indrani Bose}

\maketitle

\lyxaddress{\noindent \begin{center}
Department of Physics, Bose Institute, 93/1, A. P. C. Road, Kolkata
700009.
\par\end{center}}
\begin{abstract}
The jasmonate (JA) signaling pathway in plants is activated as defense
response to a number of stresses like attacks by pests or pathogens
and wounding by animals. Some recent experiments provide significant
new knowledge on the molecular detail and connectivity of the pathway.
The pathway has two major components in the form of feedback loops,
one negative and the other positive. We construct a minimal mathematical
model, incorporating the feedback loops, to study the dynamics of
the JA signaling pathway. The model exhibits transient gene expression
activity in the form of JA pulses in agreement with experimental observations.
The dependence of the pulse amplitude, duration and peak time on the
key parameters of the model is determined computationally. The deterministic
and stochastic aspects of the pathway dynamics are investigated using
both the full mathematical model as well as a reduced version of it.
We also compare the mechanism of pulse formation with the known mechanisms
of pulse generation in some bacterial and viral systems.
\end{abstract}

\section{Introduction}

Plants are frequently subjected to a number of stresses like attacks
by pests or pathogens and wounding by animals which cause damage to
the plant cells. In higher plants, the response to damages is mediated
via the jasmonate (JA) signaling pathway \cite{key-1,key-2,key-3,key-4}.
On activation of the pathway, there is a rapid but transient accumulation
of JAs in the infected/wounded cells. This leads to the expression
of JA- responsive genes involved in defense-related processes. More
generally, JAs are produced in response to a number of biotic and
abiotic stresses and also play active roles in regulating developmental
processes. The JA signaling pathway is an integral component of a
plant's defense system but till recently there were large gaps in
our understanding of how the pathway operates. Two experimental studies
\cite{key-5,key-6}, carried out on the model plant A. thaliana, now
provide significant knowledge on the molecular details of the signaling
pathway which helps in the elucidation of its functional features. 

\begin{figure}
\begin{centering}
\includegraphics[scale=0.5]{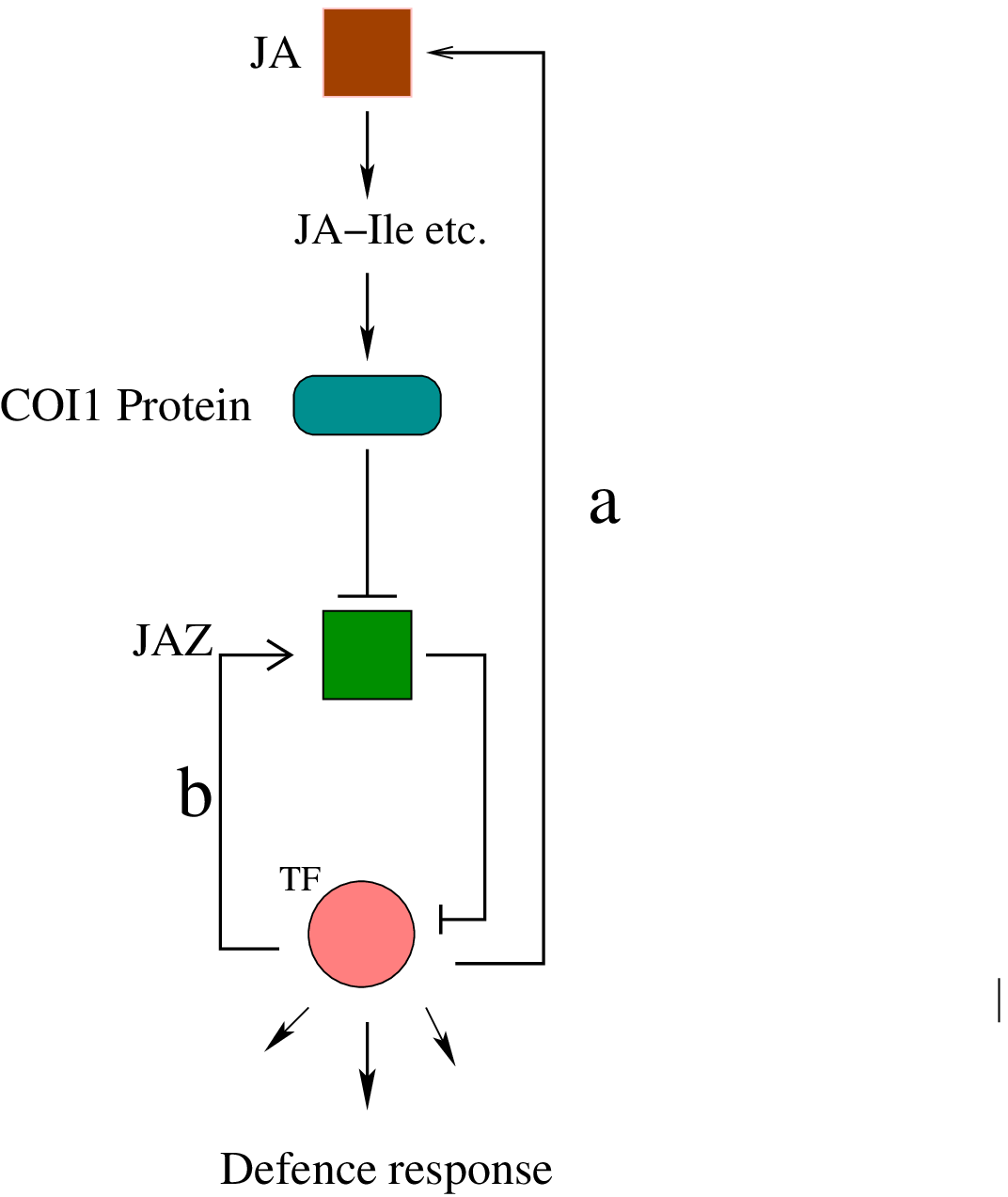}
\par\end{centering}

\caption{Core components of the JA signaling pathway \cite{key-7}. Transcription
factor (TF), MYC2, activates the synthesis of both JA and JAZ. The
arrow sign denotes activation and the hammerhead sign repression.
The function of the components is described in the text.}

\end{figure}

Figure 1 shows a schematic diagram of the core components of the JA
signaling pathway \cite{key-7}. JA responses typically involve changes
in the expressions of hundreds of genes \cite{key-1}. Transcription
factors (TFs) like MYC2 regulate the expressions of JA-responsive
genes including those promoting JA biosynthesis. Under normal circumstances,
MYC2 is non-functional due to binding by the JASMONATE-ZIM-domain
(JAZ) repressor proteins and the JA levels in cells are low. On wounding,
damage-induced signals initiate JA biosynthesis so that JA levels
are elevated in cells. This results in the formation of the jasmonate-L-isoleucine
(JA-Ile) complex which mediates the binding of CORONATINE INSENSITIVE
1(COI1) and JAZ proteins. COI1 forms part of an enzyme complex SCF$^{\text{COI1}}$
which tags the bound JAZ proteins with ubiquitin marking them for
destruction. The degradation of JAZ proteins liberates the TF MYC2
and possibly other TFs involved in JA-induced gene expression. Thus,
enhanced JA levels result in the activation of the JA signaling pathway
through the destruction of the JAZ proteins, the repressors of the
pathway. There are two primary regulatory loops \textbf{a} and \textbf{b}
in the JA signaling pathway (figure 1). Loop \textbf{b} is the newly
discovered \cite{key-5,key-6,key-7} negative feedback loop involving
MYC2 and JAZ proteins. MYC2 activates the synthesis of the JAZ proteins
whereas JAZ binds MYC2 and inhibits its activity as a TF. The synthesis
of JA is autoactivating via the positive feedback loop \textbf{a}
since a JA signal gives rise to free MYC2 which in turn activates
the genes responsible for JA biosynthesis. The combination of positive
and negative feedback loops \textbf{a} and \textbf{b} gives rise to
transient gene expression activity so that JA accumulation is in the
form of a pulse.

In this paper, we develop a minimal mathematical model incorporating
the core components of the JA signaling pathway \cite{key-7} shown
in figure 1. We specially investigate how the regulatory loops affect
the gene expression dynamics. Based on our results, we make testable
predictions on the dependence of the amplitude and duration of the
transient JA pulses on the key parameters of the model. There are
other examples of transient pulse formation in gene regulatory networks
which include the development of competence in the bacterial population
B. subtilis \cite{key-8,key-9,key-10}, infection of a host cell by
the HIV-1 virus with a subsequent choice between two distinct cell
fates, latency and lysis \cite{key-11,key-12,key-13}, and regulation
of the Salmonella pathogenecity by the PhoP/PhoQ two-component system
\cite{key-14}. The transient pulse in each case is that of a key
regulatory protein : ComK (competence development ), Tat (HIV-1 virus
) and phosphorylated PhoP ( salmonella pathogenecity ). The regulatory
proteins reach sufficiently high concentrations in the peak regions
of the pulses to activate competence development in B. subtilis, lysis
in the host cell infected by the HIV-1 virus and virulence in the
case of salmonella pathogens. In a B.subtilis population subjected
to stress, competence develops in a fraction of cells involving the
uptake of DNA from the environment and its integration into the bacterial
genome. This confers new traits on the bacteria which helps in bacterial
survival under stress. A combination of mathematical modeling and
experimental approaches provide new insight on the mechanisms of pulse
formation in the cases considered. The mechanisms differ in detail
but there is one common feature, namely, the presence of a positive
feedback loop involving the key regulatory protein.

The transient nature of a regulatory pulse is determined by two opposing
influences. In the case of B.subtilis, the competence events are generated
by two feedback loops, one positive and the other negative. Positive
feedback amplifies the ComK level whereas negative feedback brings
it down to that of the vegetative (non-competent) state. The HIV circuit
has been proposed to function like a feedback resistor \cite{key-12}
to explain transient Tat activity. Forward reactions ( e.g., acetylation
of Tat ) constituting the positive feedback loop amplify the Tat level
whereas stronger back reactions (deacetylation, degradation and unbinding
of the regulatory molecules) eventually deactivate the circuit. The
two-component PhoP/PhoQ system controlling salmonella pathogenicity
operates on similar principles \cite{key-14}. These known mechanisms
of transient pulse formation provide the motivation for identifying
the mechanism underlying transient pulse generation in the JA signaling
pathway. In section 2, we develop a minimal mathematical model of
the JA signaling pathway to study how JA pulses are formed. The dependence
of the pulse characteristics like amplitude, duration and peak time
on the key parameters of the model is computed. We further consider
a reduced version of the general model to capture the essential features
of the dynamics of pulse formation. In section 3, we investigate the
dynamics of the JA signaling pathway using stochastic approaches.
Section 4 contains concluding remarks and a comparison of the mechanisms
of pulse formation in different signaling pathways.

\section{Mathematical model of JA-Signaling Pathway}

In the deterministic approach, the dynamics of the JA-signaling pathway
are determined by a set of differential equations describing the rates
of change in the concentrations of the key molecular components in
the pathway (figure1) with respect to time. The biochemical reactions
to be considered are:

\begin{equation}
JZ+M\overset{k_{1}}{\underset{k_{2}}{\rightleftharpoons}}M_{-}JZ\end{equation}

\begin{equation}
JA+IL\overset{k_{3}}{\underset{k_{4}}{\rightleftharpoons}}JAIL\end{equation}

\begin{equation}
JZ+JAIL\overset{k_{5}}{\underset{k_{6}}{\rightleftharpoons}}JZ_{-}JAIL\overset{k_{7}}{\rightarrow}JAIL\end{equation}
 $JZ$, $JA$ and $M$ denote the molecular species JAZ, JA and MYC2.
$IL$, $M_{-}JZ$, $JAIL$ and $JZ_{-}JAIL$ represent isoleucine,
the bound complex of JAZ and MYC2, the JA-Ile complex and the complex
of JAZ with JA-Ile respectively. The same symbols represent the concentrations
of the respective molecular species. The rate constants $k_{1}$-$k_{7}$
are associated with the different forward and backward reactions.
Two other reactions which we take into account describe the degradation
of JAZ and JA. As mentioned in the Introduction, we develop a minimal
model of the JA-signaling pathway to probe the mechanism of JA pulse
formation. In the following, we point out the simplifications made
in developing the model and the justifications thereof. JA and JA-Ile
levels are known to increase dramatically ( $\sim$ 25 fold increase
within five minutes of tissue injury ) in response to both mechanical
wounding and herbivory \cite{key-37}. The remarkable speed of response
suggests that all the biosynthetic enzymes required for the production
of JA/JA-Ile are available in the resting cells. In our minimal model,
the enzymatic activity is implicit in the effective reactions considered
with the enzyme concentrations absorbed in the appropriate rate constants.
The conjugation of JA to Ile is mediated by the enzyme JASMONATE RESISTANT1
(JAR1) \cite{key-38} which is described by the effective binding
reaction in equation (2). Recent reports \cite{key-1,key-5,key-6,key-7,key-39}
have conclusively established that JA-Ile directly induces the binding
of the JAZ proteins with COI1 which results in the destruction of
the repressor proteins. These processes are combined in the effective
reactions contained in equation (3). Experimental evidence of the
direct binding between COI1 and JA-Ile has not been reported so far
\cite{key-39} though a molecular mechanism for the formation of the
bound complex of JA-Ile, COI1 and JAZ has been proposed recently \cite{key-40,key-41}.
The JAZ proteins are distinguished by the presence of a highly conserved
Jas domain which mediates protein-protein interactions with both TFs
like MYC2 and COI1 \cite{key-40}. It appears that the COI1 and MYC2
proteins compete for interaction with the Jas motif \cite{key-39}.
The presence of JA determines the outcome of the competition since
the COI1-Jas domain interaction requires the presence of JA-Ile whereas
MYC2 interacts in a hormone-independent manner.

Apart from the reactions (1)-(3), the model also takes into account
the synthesis of JAZ proteins through the TF (MYC2) regulated expression
of the JAZ genes and the biosynthesis of JA. We do not model the full
octadecanoid pathway for JA biosynthesis which is activated by damage-induced
signals \cite{key-1,key-42}. One major feature of the pathway is
that the JA biosynthesis genes are regulated by TFs like MYC2, thus
constituting a positive feedback loop. In recent studies \cite{key-43}
MYC2 has been identified as a transcriptional activator of the JA
biosynthesis gene $LOX3$. This finding matches earlier observations
\cite{key-44} that JA-induced $LOX3$ expression is reduced in the
MYC2-defective mutant $jin1.$ The differential rate equations of
the model are given by:

\begin{equation}
\frac{d(JZ)}{dt}=\frac{\beta_{1}\; M}{Km_{1}+M}+k_{2}\; M_{-}JZ-k_{1}\; JZ\;.\; M+k_{6}\; JZ_{-}JAIL-k_{5}\; JZ\;.\; JAIL-\gamma_{1\;}JZ\end{equation}

\begin{equation}
\frac{d(JA)}{dt}=\frac{\beta_{2}\; M}{Km_{2}+M}+k_{4}\; JAIL-k_{3}\; JA\;.\; IL-\gamma_{2}\; JA\end{equation}

\begin{equation}
\frac{d(M_{-}JZ)}{dt}=k_{1}\; M\;.\; JZ-k_{2}\; M_{-}JZ\end{equation}

\begin{equation}
\frac{d(JAIL)}{dt}=k_{3}\; JA\;.\; IL-k_{4}\; JAIL+(k_{6}+k_{7})\; JZ_{-}JAIL-k_{5}\; JAIL\;.\; JZ\end{equation}

\begin{equation}
\frac{d(JZ_{-}JAIL)}{dt}=k_{5}\; JZ\;.\; JAIL-(k_{6}+k_{7})\; JZ_{-}JAIL\end{equation}

The first term in equation (4) describes JAZ synthesis due to MYC2-regulated
gene expression. The $\mbox{TF}$ binding/unbinding events occur on
a time scale much faster than that of protein synthesis and degradation.
One can thus assume that steady state conditions prevail in the formation
of the $\mbox{TF}$-gene complex. The MYC2- mediated synthesis of
JA is represented as an effective process by the first term in equation
(5). The rate constants $\beta_{1}$ and $\beta_{2}$ denote the maximum
rates of JAZ and JA synthesis. The synthesis rate is half-maximal
when $Km_{1},\; Km_{2}=M$, the concentration of the regulating protein
MYC2. There are two conservation  equations.

\begin{equation}
M_{t}=M+M_{-}JZ\end{equation}

\begin{equation}
IL_{t}=Il+JAIL+JZ_{-}JAIL\end{equation}

where $M_{t}$ and $IL_{t}$ are the total concentrations of MYC2
and isoleucine respectively in free as well as bound forms. The rate
constants $\gamma_{1}$ and $\gamma_{2}$ in equations (4)-(5) are
the degradation rate constants for JAZ and JA. 

\begin{figure}
\begin{centering}
\includegraphics{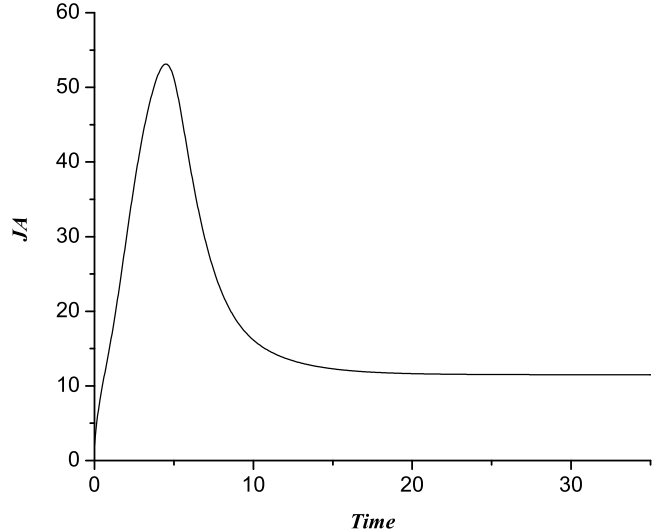}
\par\end{centering}

\caption{Transient pulse of JA versus time in hours. $JA$ has the unit of
concentration.}

\end{figure}

We now investigate whether the mathematical model, described in terms
of equations (1)-(10), can reproduce transient pulse formation consistent
with experimental observations in wounded plants \cite{key-37,key-45,key-47}.
Indeed, we find that over a wide range of parameter values JA accumulation
is in the form of a pulse. Figure 2 shows one such transient pulse
for the parameter values $\beta_{1}=30$, $\beta_{2}=30$, $k_{1}=0.2$,
$k_{2}=0.001$, $k_{3}=0.1$, $k_{4}=0.1$, $k_{5}=0.1$, $k_{6}=1.0$,
$k_{7}=1.0$, $Km_{1}=1$, $Km_{2}=1$, $\gamma_{1}=0.1$, $\gamma_{2}=0.4$
in appropriate units. The concentration of JA initially rises with
time, reaches a maximum value and then decays to reach a low steady
state value. The transient nature of the pulse is an outcome of competing
negative and positive feedback loops. An individual pulse is distinguished
by quantities like amplitude (maximal value), temporal duration and
peak time, $t_{m}$ i.e., the time at which the maximal value of the
variable being measured is achieved. The duration time $t_{D}$ is
measured as $t_{D}=t_{max2}-t_{max1}$, where $t_{max1}$ and $t_{max2}$
are respectively the time points at which the rates of increase and
decrease attain their largest values. Since positive feedback has
the effect of amplifying JA levels/activity, the amplitude of a pulse
can be increased by strengthening the positive feedback. Figure 3
shows how the JA concentration changes as a function of time for different
values of $\beta_{2}$, the maximal rate of JA synthesis. The rest
of the parameters have the same values as in the case of figure 2.
Figure 2 is obtained by solving the differential equations (4) - (8)
with the initial ( t = 0) conditions $JA$ = 2, $M$ = 100 and $Il$
= 20. The values assumed for $M_{t}$ and $IL_{t}$ are $M_{t}$=
200 and $IL_{t}$ = 20. An initial dose of JA molecules is essential
to free MYC2 so that activation of the JA signaling pathway is possible.
Wounding signals stimulate the biosynthesis of JA which provides the
initial JA input. Once free MYC2 is available, JA biosynthesis is
further stimulated via the positive feedback loop shown in figure
1. Our model calculations start with the state which already has free
MYC2 and we focus on how the activation and deactivation of the JA
pulse occurs due to the dynamics of the coupled positive and negative
feedback loops. If the initial JA and MYC2 values are increased, the
amplitude of the transient JA pulse increases. As $\beta_{2}$ becomes
larger, there is a greater accumulation of JA resulting in an increase
in the amplitude of the JA pulse. 

\begin{table}
\caption{Parameters, their meanings and defining formulae (the units are given
in brackets, $[c]$=concentration, $[t]$=time)}

\centering{}\begin{tabular}{|c|>{\raggedright}b{4in}|}
\hline 
Parameter & Meaning/defining formula\tabularnewline
\hline
\multicolumn{1}{|c|}{$\begin{array}{c}
k_{1},\; k_{2}\\
(\frac{1}{[c][t]},\frac{1}{[t]})\end{array}$} & Rate constants for the formation and dissociation respectively of
the bound complex (M\_JZ) of JAZ and MYC2 proteins (defined in equation
(1))\tabularnewline[10pt]
\hline
$\begin{array}[t]{c}
k_{3},\; k_{4}\\
(\frac{1}{[c][t]},\frac{1}{[t]})\end{array}$ & Rate constants for the formation and dissociation of the bound complex
($JAIL$) of JA with IL (defined in equation (20))\tabularnewline[10pt]
\hline
$\begin{array}{c}
k_{5},\; k_{6}\\
(\frac{1}{[c][t]},\frac{1}{[t]})\end{array}$ & Rate constants for the formation and dissociation respectively of
the complex $JZ_{-}JAIL$ of JAZ proteins with $JAIL$. (defined in
equation (3))\tabularnewline[10pt]
\hline
$\begin{array}{c}
k_{7}\\
(\frac{1}{[t]})\end{array}$ & Rate constant for ubiquitin-mediated degradation of JAZ proteins (defined
in equation (3))\tabularnewline[10pt]
\hline
$\begin{array}{c}
\gamma_{1},\;\gamma_{2}\\
(\frac{1}{[t]},\frac{1}{[t]})\end{array}$ & Degradation rate constants of JAZ proteins and JA respectively (defined
in equations (4) and (5))\tabularnewline[10pt]
\hline
$\begin{array}{c}
\beta_{1},\;\beta_{2}\\
(\frac{[c]}{[t]},\frac{[c]}{[t]})\end{array}$ & Maximum rates of JAZ and JA synthesis (defined in equations (4) and
(5))\tabularnewline[10pt]
\hline
$\begin{array}{c}
Km_{1},\; Km_{2}\\
([c],[c])\end{array}$ & Synthesis rate of JAZ and JA is half-maximal when $Km_{1}$, $Km_{2}=M$,
the concentrations of the regulatory protein $MYC2$\tabularnewline[10pt]
\hline
$\begin{array}{c}
M_{t},\; IL_{t}\\
([c],[c])\end{array}$ & Total concentrations of MYC2 and isoeucine respectively which are
defined in equations (9) and (10)\tabularnewline[10pt]
\hline
\end{tabular}
\end{table}

\begin{figure}
\begin{centering}
\includegraphics{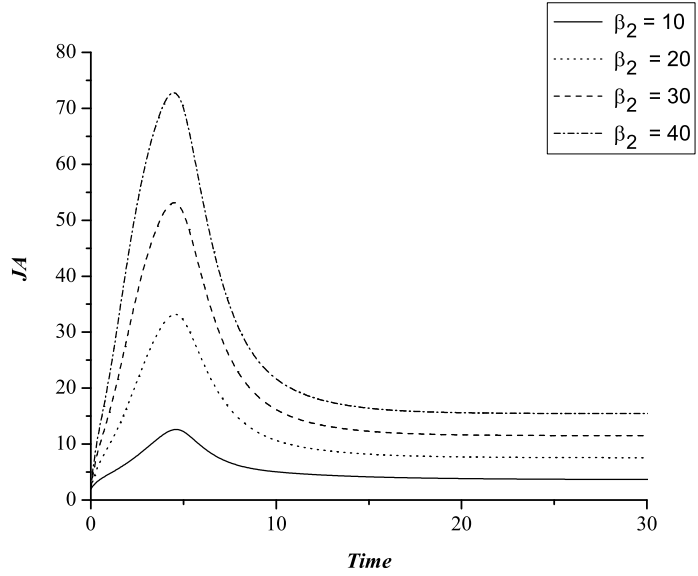}
\par\end{centering}

\caption{Variation of JA versus time in hours for values of $\beta_{2}=10,\;20\;,30$
and $40$. The peak height of the pulse increases for increasing values
of $\beta_{2}$. $JA$ has the unit of concentration.}

\end{figure}

Figure 4 shows the variation of JA concentration versus time for different
values of $\beta_{1}$, the maximal rate of JAZ synthesis. The other
parameter values are the same as in the case of figure 2. With increasing
values of $\beta_{1}$, the amount of JAZ proteins goes up and there
is more repression of MYC2 activity. This leads to a reduction in
the duration of the JA pulse. The change in the amplitude is, however,
not as prominent as that when $\beta_{2}$ is varied. This is more
evident in figures 5(a) and (b) in which the amplitude of the JA pulse
is plotted against $\beta_{1}$ and $\beta_{2}$, respectively keeping
the other parameter values the same as in the case of figure 2. Figures
5(c) and (d) exhibit the variation of the peak time $t_{m}$ versus
$\beta_{1}$ and $\beta_{2}$ respectively. The magnitude of $t_{m}$
decreases considerably as $\beta_{1}$ is increased. The dependence
of $t_{m}$ on $\beta_{2}$ is, however, insignificant. Figures 5
(e) and (f) show a plot of $t_{D}$, the duration of the JA pulse,
versus $\beta_{1}$ and $\beta_{2}$. The dependence of $t_{D}$ on
$\beta_{2}$ is negligible. Figure 6 shows the variation of JA concentration
versus time as the degradation rate constant $\gamma_{2}$ is changed.
Similar figures may be obtained by varying the other parameters of
the model. For the set of parameter values considered, the most prominent
changes are obtained by varying the maximum synthesis rates $\beta_{1}$
and $\beta_{2}$. Figures 3-6 have been obtained with the total concentrations
fixed at the values $M_{t}=200$, and $IL_{t}=20$ in appropriate
units. Table 1 lists all the parameters of the dynamical model, their
description as well as units. The different parameter units are expressed
in terms of concentration ({[}c{]}) and time ({[}t{]}) units. There
are presently no methods available to record the cellular concentrations
of JA, JA-Ile etc., the quantities measured are amounts per gram fresh
weight. In our model study, the JA amounts have the units of concentration.

\begin{figure}
\begin{centering}
\includegraphics{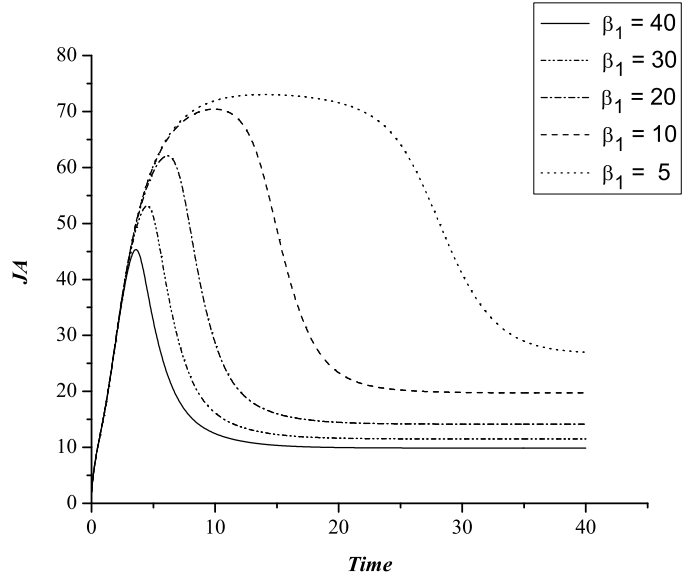}
\par\end{centering}

\caption{Variation of $JA$ versus time for different values of $\beta_{1}=5,\;10,\;20\;,30$
and $40$.}

\end{figure}

\begin{figure}
\begin{centering}
\includegraphics[scale=0.8]{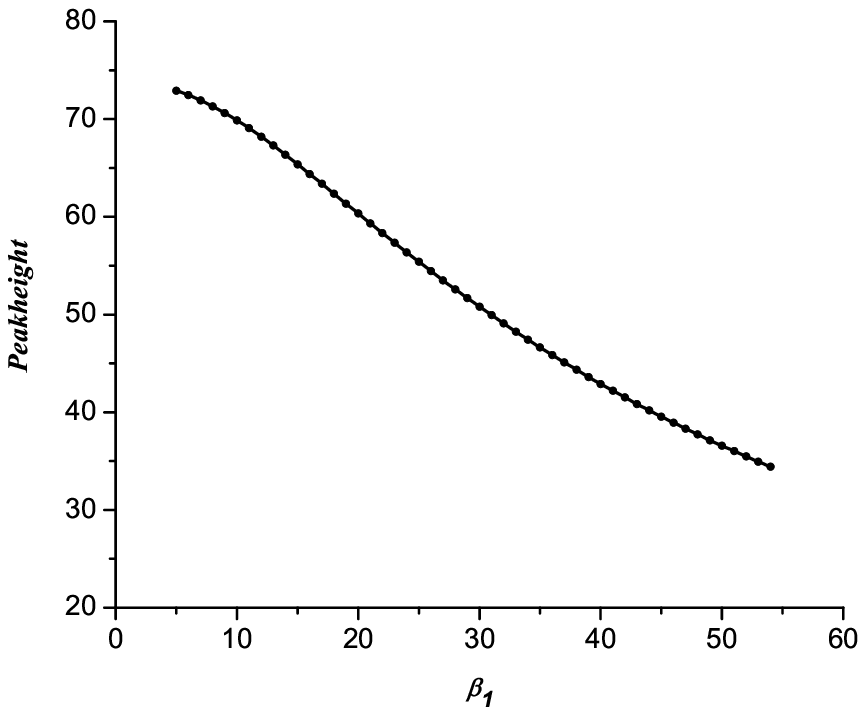}\includegraphics[scale=0.8]{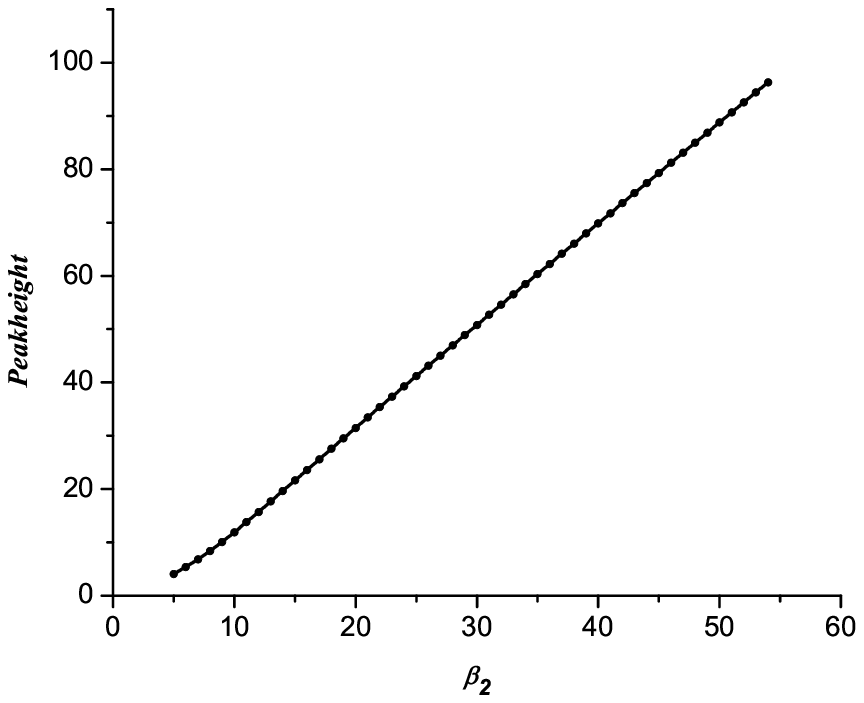}
\par\end{centering}

\begin{centering}
\includegraphics[scale=0.8]{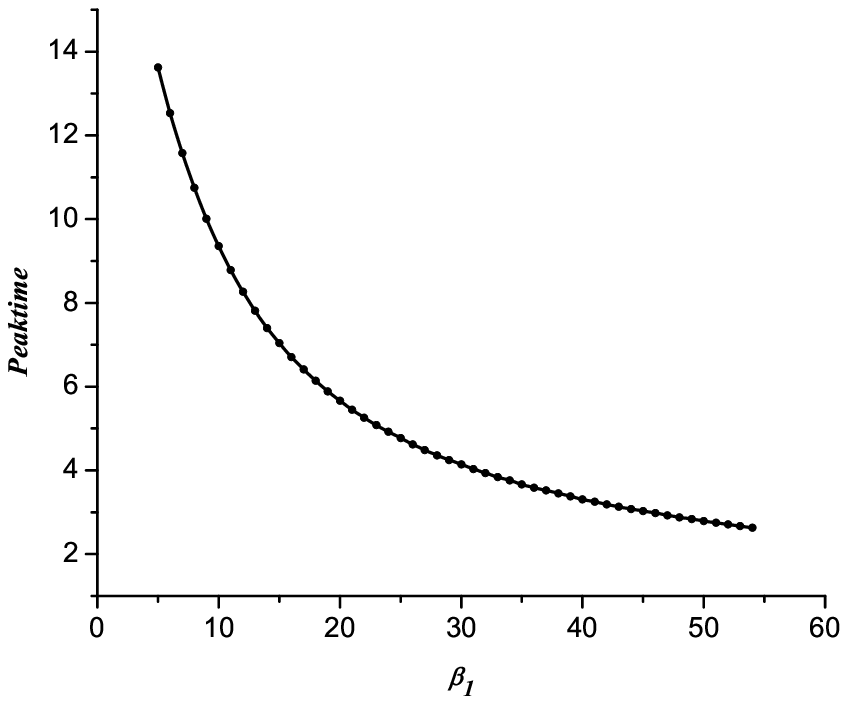}\includegraphics[scale=0.8]{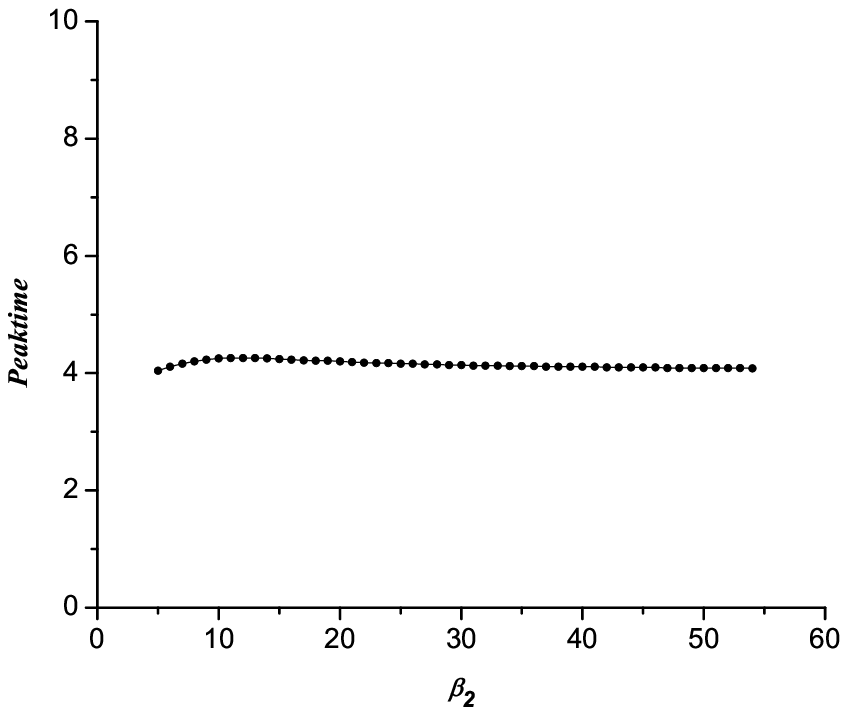}
\par\end{centering}

\begin{centering}
\includegraphics[scale=0.8]{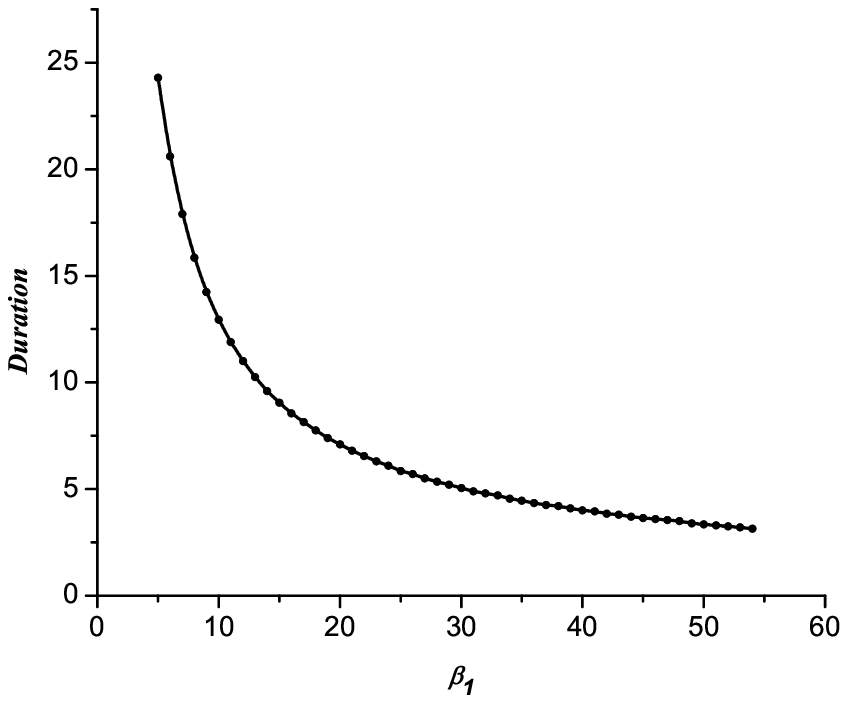}\includegraphics[scale=0.8]{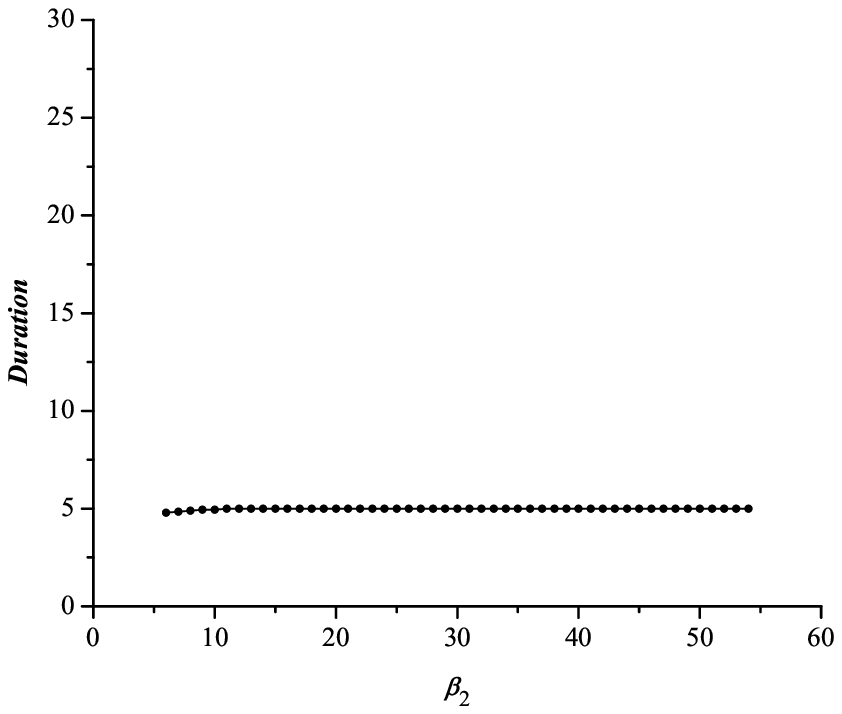}
\par\end{centering}

\caption{Amplitude of JA pulse versus (a) $\beta_{1}$ and (b) $\beta_{2}$
; Peak time $t_{m}$ in hours of the JA pulse versus (c) $\beta_{1}$
and (d) $\beta_{2}$; Duration $t_{D}$ in hours of JA pulse versus
(e) $\beta_{1}$ and $\beta_{2}$.}

\end{figure}

To analyse the features of the model better, we reduce it to a two-variable
model described by rate equations which are in dimensionless form.The
reduction rests on the fact that the molecular complexes $M_{-}JZ$,
$JAIL$ and $JZ_{-}JAIL$ attain their steady state concentrations
on timescales which are comparably shorter than those over which the
JA and JAZ reach their steady state concentrations. Therefore, one
can take the time derivatives in equations (6)-(8) to be zero so that
the complexes $M_{-}JZ$, $JAIL$ and $JZ_{-}JAIL$, attain their
steady state concentrations. The reduced model is described by the
rate equations \begin{equation}
\frac{d(JZ)}{dt}=\frac{\beta_{1\;}M}{Km_{1}+M}-\delta_{JZ}\frac{JA\;.\; JZ}{JA+K1+\frac{JZ}{\Gamma}\;.\; JA}-\gamma_{1\;}JZ\end{equation}

\begin{equation}
\frac{d(JA)}{dt}=\frac{\beta_{2}\; M}{Km_{2}+M}-\gamma_{2}\; JA\end{equation}

with $\Gamma=\frac{k_{6}+k_{7}}{k_{5}}$, $K1=\frac{k_{4}}{k_{3}}$
and $\delta_{JZ}=\frac{k_{7\;}IL_{t}}{\Gamma}$. Also, $M_{t}=M(1+\frac{k_{1}}{k_{2}}\; JZ)$
and $IL_{t}=IL(1+\frac{k3}{k4}JA(1+\frac{JZ}{\Gamma}))$. Rescaling
time with $\delta_{JZ}$ and the concentrations JZ and JA with $\Gamma$,
we obtain the dimensionless rate equations:

\begin{equation}
\frac{d(JZ)}{dt}=\frac{b_{1}}{kn_{1}(1+K^{'}\; JZ)+1}-\frac{JA\;.\; JZ}{JA+K2+JA\;.\; JZ}-\delta_{1}\; JZ\end{equation}

\begin{equation}
\frac{d(JZ)}{dt}=\frac{b_{2}}{kn_{2}(1+K^{'}\; JZ)}-\delta_{2}\: JA\end{equation}

where $t$, $JZ$ and $JA$ are now dimensionless variables with the
dimensionless parameters shown in Table 2.

\begin{table}
\caption{Rescaled parameters and their defintions}

\begin{doublespace}
\centering{}\begin{tabular}{|c|c|c|}
\hline 
Parameter & Definition & Abbreviation\tabularnewline
\hline
$b_{1}$ & $\frac{\beta_{1}}{\delta_{JZ\;}\Gamma}$ & $\delta_{JZ}=\frac{k_{7}\; IL_{t}}{\Gamma}$\tabularnewline
$b_{2}$ & $\frac{\beta_{2}}{\delta_{JZ}\;\Gamma}$ & $\Gamma=\frac{k_{6}+k_{7}}{k_{5}}$\tabularnewline
$kn_{1}$ & $\frac{Km_{1}}{M_{t}}$ & $M_{t}=M(1+\frac{k_{1}}{k_{2}}JZ)$\tabularnewline
$kn_{2}$ & $\frac{Km_{2}}{M_{t}}$ & $IL_{t}=IL(1+\frac{k_{3}}{k_{4}}JA(1+\frac{JZ}{\Gamma}))$\tabularnewline
$K^{'}$ & $\Gamma\frac{k_{1}}{k_{2}}$ & \tabularnewline
$K2$ & $\frac{K1}{\Gamma}$ & \tabularnewline
$\delta_{1}$ & $\frac{\gamma_{1}}{\delta_{JZ}}$ & \tabularnewline
$\delta_{2}$ & $\frac{\gamma_{2}}{\delta_{JZ}}$ & \tabularnewline
$\alpha_{1}$ & $\frac{b_{1}}{kn_{1}+1}$ & \tabularnewline
$\alpha_{2}$ & $\frac{b_{2}}{kn_{2}+1}$ & \tabularnewline
$\mu_{1}$ & $\frac{kn_{1\;}K^{'}}{(kn_{1}+1)}$ & \tabularnewline
$\mu_{2}$ & $\frac{kn_{2\;}K^{'}}{(kn_{2}+1)}$ & \tabularnewline
\hline
\end{tabular}\end{doublespace}

\end{table}

Equations (13)-(14) can be further simplified to

\begin{equation}
\frac{d(JZ)}{dt}=\frac{\alpha_{1}}{1+\mu_{1}\; JZ}-\frac{JA\;.\; JZ}{JA+K2+JA\;.\; JZ}-\delta_{1}\; JZ\end{equation}

\begin{equation}
\frac{d(JA)}{dt}=\frac{\alpha_{2}}{1+\mu_{2\;}JZ}-\delta_{2}\; JA\end{equation}

The parameters $\alpha_{2}$, $\alpha_{2}$, $\mu_{1}$ and $\mu_{2}$
are defined in Table 2. We first consider the case $\delta_{1}=0$
as degradation of the JAZ proteins mostly occurs via the second term
in Eq.(13) describing ubiquitin-mediated degradation on the binding
of the SCF$^{\text{COI1}}$ complex to JAZ proteins. The steady state
of the dynamical system is obtained by putting the temporal rates
of change, $\frac{d(JZ)}{dt}$ and $\frac{d(JA)}{dt}$, to be zero.
There is only one physical steady state solution given by 

\begin{equation}
(JZ)_{s}=\frac{A+\sqrt{A^{2}+4\:\alpha_{1}\:\alpha_{3}\:\mu_{1\:}(\alpha_{3}\,+\, K2)}}{2\:\alpha_{3}\:\mu_{1}}\qquad A=\alpha_{1\:}\alpha_{3}+\alpha_{1\:}K2\:\mu_{2}-\alpha_{3}\end{equation}

\begin{equation}
(JA)_{s}=\frac{\alpha_{3}}{1+\mu_{2\:}(JZ)_{s}}\quad\alpha_{3}=\frac{\alpha_{2}}{\delta_{2}}\end{equation}

\begin{figure}
\begin{centering}
\includegraphics{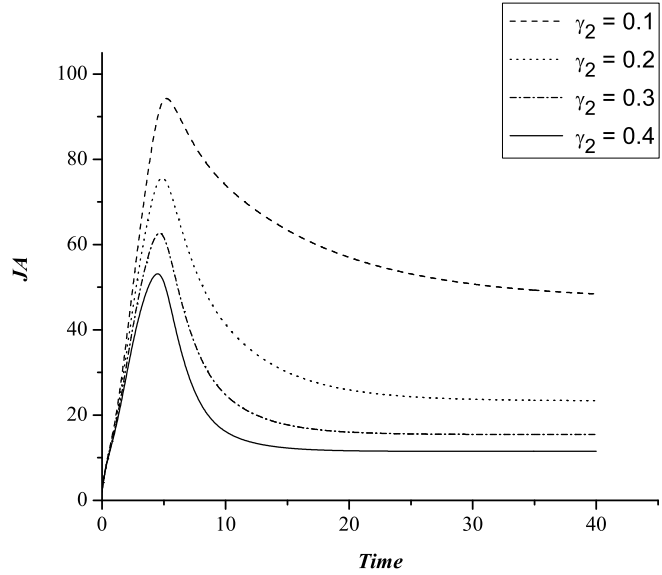}
\par\end{centering}

\caption{Variation of $JA$ versus time in hours for different values $\gamma_{2}=0.1\;,0.2\;,0.3$
and $0.4$ of the degradation rate constant. The amplitude of the
JA pulse decreases for increasing value of $\gamma_{2}$. $JA$ has
the unit of concentration.}

\end{figure}

The steady state is found to be stable over a wide range of parameter
values. When $\delta_{1}$ is $\neq$ $0$ in equation (15), again
over a wide range of parameter values there is only one physical steady
state solution which is furthermore stable. Figure 7 shows the phase
portrait obtained from equations (15) and (16) with the parameter
values $\alpha_{1}=2.5$, $\alpha_{3}=50$, $\mu_{1}=50$, $\mu_{2}=10$,
$K2=1000$, and $\delta_{1}=0.015$. The nullclines, obtained by putting
$\frac{d(JA)}{dt}=0$ and $\frac{d(JZ)}{dt}=0$, intersect at a single
point, the so-called fixed point of the dynamics. Two trajectories
(shown by dotted lines) with two different initial conditions converge
to the fixed point in the course of time. Each trajectory depicts
the values of $JA$ and $JZ$ at different time points obtained by
solving equations (15) and (16). Figure 8 shows the magnitude of $JA$
versus time for the same parameter values. The transient pulse reaches
a peak value due to an initial surge and the system finally settles
down into a stable steady state with the value of $JA$ much lower
than the peak value.

\begin{figure}
\begin{centering}
\includegraphics{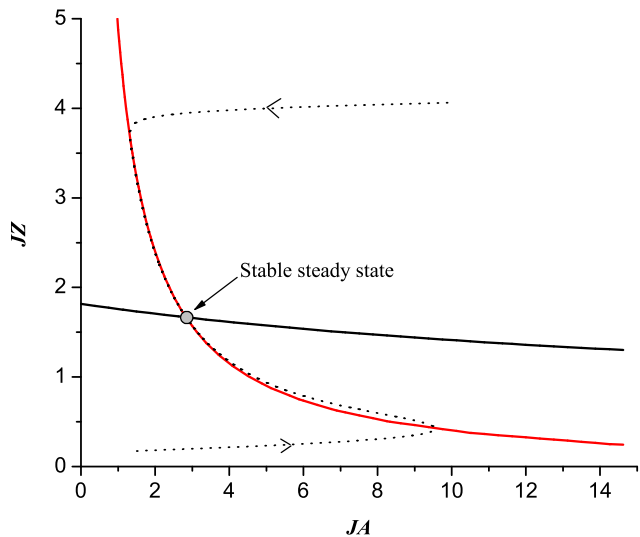}
\par\end{centering}

\caption{Phase portrait defined by equations (15) and (16). The solid lines
represent the nullclines($\frac{d(JA)}{dt}=0$(red), $\frac{d(JZ)}{dt}=0$(black))
intersecting at one fixed point. The fixed point describes a stable
steady state. Two typical trajectories (dotted lines) are shown with
arrow directions denoting increasing time.}

\end{figure}

\begin{figure}
\begin{centering}
\includegraphics{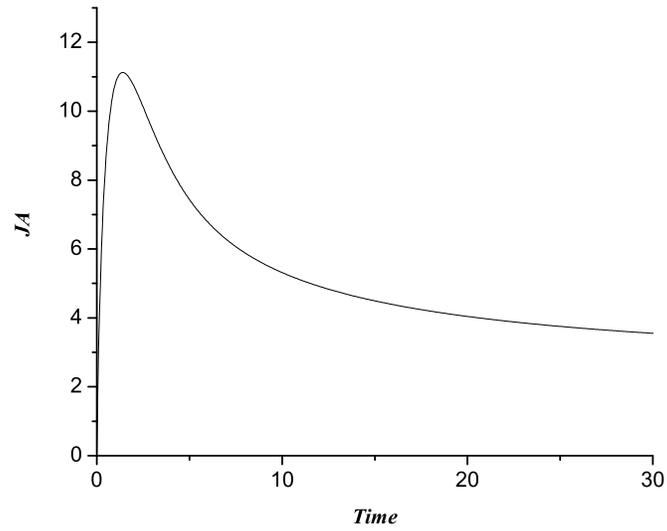}
\par\end{centering}

\caption{Transient pulse of JA versus time in the reduced model defined by
equations (15) and (16). $JA$ and time are dimensionless in the reduced
model. }

\end{figure}

\begin{figure}
\begin{centering}
\includegraphics{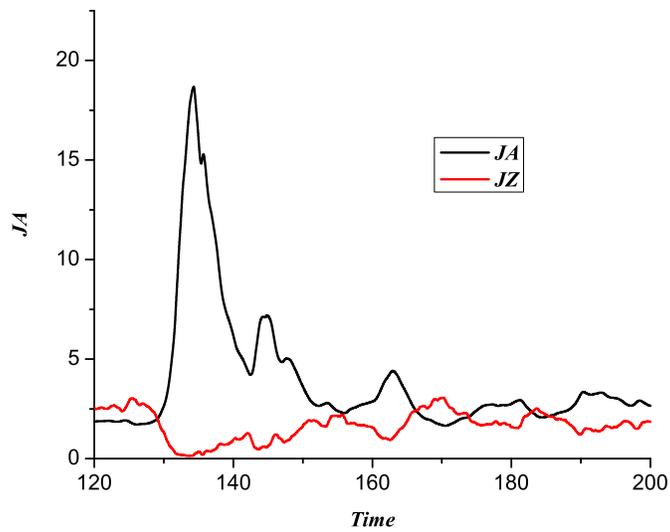}
\par\end{centering}

\caption{Variation of $JA$ and $JZ$ versus time in dimensionless units.}

\end{figure}

\begin{figure}
\begin{centering}
\includegraphics{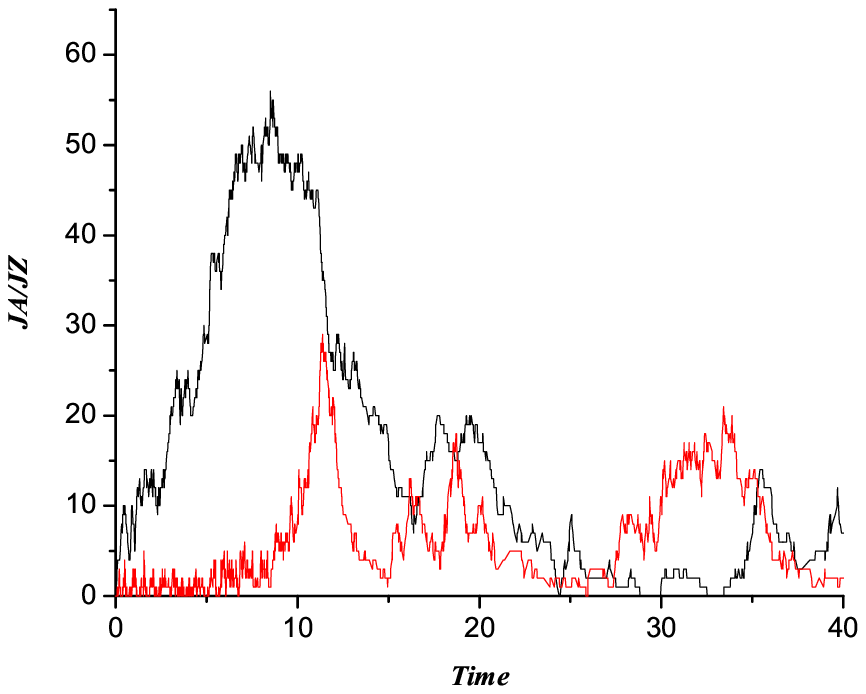}
\par\end{centering}

\caption{Number of JA (black curve) and JAZ (red curve) molecules versus time
in hours. The values of the stochastic rate constants  are mentioned
in the text.}

\end{figure}

\section{Stochastic Dynamics}

The single steady state characterized by a low value of \textit{JA}
is stable under small perturbations. Under sufficiently strong perturbations,
a large-amplitude JA pulse is excited with eventual return to the
stable steady state. The random nature of the biochemical events involved
in gene expression gives rise to fluctuations in protein levels \cite{key-15,key-16}
which act as perturbations. The simplest way to take the effect of
fluctuations into account is to incorporate an additive noise term
in either of the dynamical equations (15) and (16). The noise term
further incorporates the effects of transient perturbations originating
in wounding signals. The most prominent effect is obtained with the
noise term added to equation (15). A possible reason for this is that
JAZ represses the synthesis of both JA and JAZ so that fluctuations
in the JAZ levels affect both the JA and JAZ levels. The stochastic
dynamics are described by

\begin{equation}
\frac{d(JZ)}{dt}=\frac{\alpha_{1}}{1+\mu_{1}\; JZ}-\frac{JA\;.\; JZ}{JA+K2+JA\;.\; JZ}-\delta_{1}\; JZ+\xi(t)\end{equation}

\begin{equation}
\frac{d(JA)}{dt}=\frac{\alpha_{2}}{1+\mu_{2}\; JZ}-\delta_{2}\; JA\end{equation}

The noise term $\xi(t)$ approximates the actual fluctuations which
occur in the system. We assume the noise to be of the Ornstein-Uhlenbeck(OU)
type (coloured noise ) \cite{key-17,key-18} as recent studies show
that OU noise plays a significant role in gene expression dynamics
\cite{key-9,key-19}. The noise variable has zero mean and an exponentially
decaying correlation in time, i.e., 

\begin{equation}
\left\langle \xi(t)\right\rangle =0\end{equation}

\begin{equation}
\left\langle \xi(t)\;\xi(t^{'})\right\rangle =D\;\lambda\; exp(-\lambda\mid t-t^{'}\mid)\end{equation}

where D is the noise strength and $\lambda$ the inverse of the correlation
time $\tau$. We use the numerical simulation algorithm developed
for the OU process \cite{key-20,key-21} to obtain solutions of the
stochastic differential equations (19) and (20). The major steps of
the algorithm are as follows. Let $x=JA$ and $y=JZ$. Knowing the
values of $x$ and $y$ at time $t$, the values at time $t+\triangle t$
( $\triangle t$ is small) are 

\begin{equation}
x(t+\triangle t)=x(t)+p\;\triangle t\end{equation}

\begin{equation}
y(t+\triangle t)=y(t)+q\;\triangle t\end{equation}

where 

\begin{equation}
p=f(x,y)+\xi(t)\end{equation}

\begin{equation}
q=g(x,y)\end{equation}

The functions $f(x,y)$ and $g(x,y)$ are, from equation (19) and
(20),

\begin{equation}
f(x,y)=\frac{\alpha_{1}}{1+\mu_{1}\; y}-\frac{x\;.\; y}{x+K2+x\;.\; y}-\delta_{1}\; y\end{equation}

\begin{equation}
g(x,y)=\frac{\alpha_{2}}{1+\mu_{2}\; y}-x\end{equation}

The noise variable $\xi(t+\triangle t)$ is 

\begin{equation}
\xi(t+\triangle t)=\xi(t)\mu+\sigma_{x}\; n_{1}\end{equation}

with 

\begin{equation}
\mu=e^{-\lambda\;\triangle t}\end{equation}
\begin{equation}
\sigma_{x}^{2}=D\;\lambda(1-\mu^{2})\end{equation}

and $n_{1}$ is a normal random variable with mean zero and variance
1. Through iterative computation, using equations (23)-(31), the values
of $x$ and $y$ at successive time steps $\triangle t$ can be determined.
Figure 9  shows the time traces of $x(JA)$ and $y(JZ)$ for the parameter
values the same as in the cases of figures 7 and 8. The initial values
of $x$ and $y$ are zero, $\frac{1}{\lambda}=\tau=0.5$ and $D=0.0625$.
The incremental time step $\triangle t=0.01$. Figure 9  shows the
anticorrelation between $JA$ and $JZ$ values, the $JZ$ value dips
when $JA$ rises up to the peak pulse value. Fluctuations of moderate
strength in $JZ$ can give rise to transient JA pulses. With increased
noise strength, transient pulses are more readily generated.

We further probe the stochastic dynamics of the full model with the
help of computer simulation based on the Gillespie algorithm \cite{key-22}.
The reactions are 15 in number and are given by 

\begin{equation}
M+JZ\rightarrow M_{-}JZ\end{equation}

\begin{equation}
M_{-}JZ\rightarrow M+JZ\end{equation}

\begin{equation}
JA+IL\rightarrow JAIL\end{equation}

\begin{equation}
JAIL\rightarrow JA+IL\end{equation}

\begin{equation}
JZ+JAIL\rightarrow JA_{-}JAIL\end{equation}

\begin{equation}
JZ_{-}JAIL\rightarrow JZ+JAIL\end{equation}

\begin{equation}
JZ_{-}JAIL\rightarrow JAIL\end{equation}

\begin{equation}
G1^{*}\rightarrow JZ\end{equation}

\begin{equation}
G2^{*}\rightarrow JA\end{equation}

\begin{equation}
JZ\rightarrow\phi\end{equation}

\begin{equation}
JA\rightarrow\phi\end{equation}

\begin{equation}
G1+M\rightarrow G1^{*}\end{equation}

\begin{equation}
G1^{*}\rightarrow G1+M\end{equation}

\begin{equation}
G2+M\rightarrow G2^{*}\end{equation}

\begin{equation}
G2^{*}\rightarrow G2+M\end{equation}

The different symbols are explained in section 2. $\phi$ denotes
the proteins degradation product. $G1$ ($G2$) and $G1^{*}$ ($G2^{*}$)
are the inactive and active states respectively of the gene synthesizing
JAZ proteins (JA). The active state of the gene is attained on the
binding of the transcription factor MYC2 to the regulatory region
of the gene. For simplicity, only one JA biosynthesis gene is assumed.
The stochastic rate constants, associated with the reactions, are
$C(i),\; i=1,.....15$ in appropriate units. Figure 10 shows the variation
of the number of JA (black curve) and JAZ (red curve) molecules as
a function of time. The values of the stochastic rate constants are
$C(1)=0.2$, $C(2)=0.001$, $C(3)=0.1$, $C(4)=0.1$, $C(5)=0.1$,
$C(6)=1.0$, $C(7)=1.0$, $C(8)=12.0$, $C(9)=10.0$, $C(10)=0.1$,
$C(11)=0.4$, $C(12)=1.0$, $C(13)=1.0$, $C(14)=1.0$ and $C(15)=1.0$.
The stochastic time evolution of $JA$ shows a transient pulse of
large amplitude followed by a series of small amplitude transient
pulses. As in the case of the reduced model, the JA and JZ values
are anticorrelated.

\section{Conclusion and Outlook}

The JA signaling pathway is activated in defending plants against
attacks by pathogens and wounding by animals. Recent experiments \cite{key-4,key-5,key-6}
provide important new knowledge on the molecular connectivity of the
pathway. The observations suggest a model in which two feedback loops,
one positive and the other negative, control the dynamics of JA response
in the form of a transient pulse. In this paper, we propose a minimal
mathematical model which captures the essential features of the JA
signaling pathway. To our knowledge, no other mathematical model of
the dynamics of the coupled feedback loops in the JA signaling pathway
has been proposed so far. The model demonstrates the formation of
the transient JA pulse for different parameter values. The pulse is
an outcome of two opposing influences mediated via the positive and
negative feedback loops.We compute the dependence of the pulse amplitude,
duration and peak time on the key parameters of the theoretical model.
The full mathematical model, described in terms of equations (4)-(8),
can be reduced to a two-variable model (equations (15) and (16)) for
which the physical steady state solution is obtained analytically.
There is only one such solution, corresponding to low JA levels, which
is stable in an extended region of the parameter space. A similar
conclusion holds true for the full dynamical model. We explore the
dynamics of both the full and reduced models using deterministic as
well as stochastic formalisms. The TF MYC2 activates the expression
of both the JA biosynthesis and JAZ genes. In normal circumstances,
the expression activity is low as JAZ proteins repress MYC2 by forming
bound complexes with the TF proteins. For the JA pulse to be initiated,
free MYC2 is required. Once the JA synthesis is initiated, the positive
feedback loop ramps up further JA production. The negative feedback
loop acts against the enhancement so that the JA level attains a steady
lower value after a certain time interval. JA-mediated destruction
of JAZ repressors frees MYC2 which in turn activates JAZ gene expression.
The production of JAZ proteins completes the negative feedback loop
through the repression of MYC2 activity (figure 1). In the deterministic
formalism, the transient JA pulse is obtained when a certain amount
of free MYC2 is available at time $t=0$. The availability of free
MYC2 is JA-dependent. We do not model the explicit steps of JA biosynthesis
and the synthesis of MYC2. We study the dynamics of the JA signaling
pathway with free MYC2 as input. The central role of MYC2 in the model
is dictated by available experimental evidence but there could be
other TFs playing similar roles (see figure 5(b) of \cite{key-28}).
The prominent role of MYC2 could be tested experimentally by adding
free MYC2 to plant leaves and checking whether transient JA pulses
are obtained even in the absence of wounding. In the stochastic case,
fluctuations in the numbers of JA and JAZ proteins can activate pulse
formation. This is clearly understood in the framework of the reduced
model. As figure 9 distinctly shows, the fluctuations in JA and JAZ
numbers are anticorrelated. A temporary dip in the JAZ levels results
in the increased availability of free MYC2 and the subsequent generation
of a JA pulse. Single cell experiments on a variety of organisms have
established the stochastic nature of gene expression \cite{key-15,key-16}
and it is now common knowledge that protein fluctuations have a considerable
effect on cellular processes.

The operating principle behind transient gene expression activity
in the JA signaling pathway is based on a combination of positive
and negative feedback loops and the availability of free MYC2 to transactivate
the expression of the JA biosynthesis and JAZ genes. Due to the transient
nature of the JA pulse, individual cells attain a state in which elevated
levels of JA induce the expression of target genes involved in defense
response and after a time interval return  to the quiescent state.
In B. subtilis, a core module consisting of positive and negative
feedback loops can explain transient gene expression activity resulting
in entry into the competent state and subsequent exit from it \cite{key-9}.
The dynamics, however, have features different from those in the case
of the JA signaling pathway. The ComK proteins autoactivate their
own production in a cooperative fashion, i.e., bound complexes of
ComK activate the initiation of transcription of the $comK$ gene.
The analysis of the dynamics reveals three physical fixed points:
a stable fixed point at low ComK concentration ( the vegetative state),
one unstable saddle fixed point at intermediate ComK and an unstable
spiral at high ComK ( the competent state) \cite{key-9}. The system
exhibits excitable dynamics in which relatively small perturbations
of the low ComK vegetative state can initiate long excursions in phase
space around the unstable spiral node, resulting in transient activation
of the competent state. Since the intermediate and high ComK fixed
points are unstable, the system ultimately returns to the stable vegetative
state. In contrast, the dynamics of JA pulse formation involve only
one fixed point. A similar situation prevails in the case of the HIV
regulatory circuit. A minimal model of the circuit dynamics predicts
a single fixed point which is stable provided a specific condition
is satisfied \cite{key-12}. In this case, the long-lived pulse of
Tat proteins results from transient positive feedback amplification
and deactivation via stronger back reactions like deacetylation of
the Tat protein and unbinding of the regulatory molecules. The back
reactions opposing the action of the positive feedback loop have an
effective influence similar to that of a negative feedback loop. The
operating principle in the HIV circuit is that of feedback resistor,
analogous to the presence of a dissipative resistor in an electrical
feedback circuit. The dissipative resistor stabilizes the low expression
state of a positive feedback loop.

In B. subtilis, fluctuations in ComK levels trigger competence development
in a fraction of cells, demonstrated explicitly in the stochastic
dynamics of the minimal model describing competence development \cite{key-9}.
There is now experimental validation of the proposal that noise in
gene expression determines cell fate in B. subtilis \cite{key-10}.
In the Tat circuit, stochastic molecular fluctuations intrinsic to
gene expression can activate transient Tat pulses. The feedback resistor
model, however, shows that the off state is quite stable and more
robust to noise than cooperative positive feedback loops. The inherent
noise buffering is due to the presence of \ {}``dissipative resistors''
in the transactivation circuit \cite{key-12}. The phenotypic diversity
observed in isogenic HIV-1 virus populations is ascribed to desynchronized
Tat pulses in individual cells due to stochastic gene expression.
In the case of the JA signaling pathway, noise-induced activation
of the JA pulse is possible (figures 9 and 10) but the off state is
found to be quite stable to fluctuations in the stochastic simulation
of the reduced ( figure 9) and full (figure 10) models. The negative
feedback, analogous to the feedback resistor, enhances the off-sate
stability in this case. 

An interesting aspect of JA-response not addressed in the present
study relates to memory formation of past wounding events or attacks
priming plants for a more effective JA-mediated defense response to
subsequent attacks \cite{key-23,key-24}. Evidence of memory formation
could lie in the modification of (1) the peak time of JA accumulation,
(2) amplitude of JA pulse or (3) baseline of accumulated JA due to
discrete JA bursts elicited by repeated attacks \cite{key-23}. The
negative feedback loop \textbf{b} in the JA-signaling network constitutes
a well-known motif in gene regulatory networks. A negative feedback
loop by itself may give rise to oscillatory dynamics or alternatively
pulsed protein production \cite{key-25,key-26}. The $p53-MDM2$ loop
provides a well-known example of negative feedback \cite{key-25}.
The loop is functionally active in response to DNA damage. Single
cell experiments show that on DNA damage by $\gamma$-irradiation
a series of undamped p53 pulses of fixed amplitude and duration is
generated. The amount of $\gamma-$irradiation determines the number
of pulses but has no effect on the amplitude and duration of the pulses.
The functional response of the MYC2-JAZ negative feedback loop in
the JA signaling pathway to wounding signals of varying magnitude
is worth exploring at the single cell level to gain insight on key
dynamical features. It is generally difficult to link circuit dynamics
with circuit structure. The identification of feedback loops in the
JA signaling pathway allows for the construction of a minimal mathematical
model which correctly reproduces the formation of JA pulses on activation.
The deterministic and stochastic origins of pulse formation are obtained
through computational analysis. The dependence of key pulse parameters
on the kinetic rate constants is demonstrated quantitatively. Further
experiments are needed to validate the theoretical predictions as
well as to elucidate the operating principle of the JA signaling pathway
at a microscopic level. Comparative studies of operating principles
controlling transient gene expression activity in diverse organisms
are expected to provide insight on the common mechanisms organisms
living in a complex environment adopt for responding to stress signals.

\pagebreak{}

Figure Captions

1. Core components of the JA signaling pathway \cite{key-7}. Transcription
factor (TF), MYC2, activates the synthesis of both JA and JAZ. The
arrow sign denotes activation and the hammerhead sign repression.
The function of the components is described in the text.

2. Transient pulse of JA versus time in hours. $JA$ has the unit
of concentration.

3. Variation of JA versus time in hours for values of $\beta_{2}=10,\;20\;,30$
and $40$. The peak height of the pulse increases for increasing values
of $\beta_{2}$. $JA$ has the unit of concentration.

4. Variation of $JA$ versus time for different values of $\beta_{1}=5,\;10,\;20\;,30$
and $40$.

5. Amplitude of JA pulse versus (a) $\beta_{1}$ and (b) $\beta_{2}$
; Peak time $t_{m}$ in hours of the JA pulse versus (c) $\beta_{1}$
and (d) $\beta_{2}$; Duration $t_{D}$ in hours of JA pulse versus
(e) $\beta_{1}$ and $\beta_{2}$.

6. Variation of $JA$ versus time in hours for different values $\gamma_{2}=0.1\;,0.2\;,0.3$
and $0.4$ of the degradation rate constant. The amplitude of the
JA pulse decreases for increasing value of $\gamma_{2}$. $JA$ has
the unit of concentration.

7. Phase portrait defined by equations (15) and (16). The solid lines
represent the nullclines($\frac{d(JA)}{dt}=0$(red), $\frac{d(JZ)}{dt}=0$(black))
intersecting at one fixed point. The fixed point describes a stable
steady state. Two typical trajectories (dotted lines) are shown with
arrow directions denoting increasing time.

8. Transient pulse of JA versus time in the reduced model defined
by equations (15) and (16). $JA$ and time are dimensionless in the
reduced model. 

9. Variation of $JA$ and $JZ$ versus time in dimensionless units.

10. Number of JA (black curve) and JAZ (red curve) molecules versus
time in hours. The values of the stochastic rate constants are mentioned
in the text.
\end{document}